\documentstyle[12pt]{article}

\begin{document}
\title{Numerical study of pattern formation following a convective
instability in non-Boussinesq fluids}
\author{Hao-wen Xi\\
Department of Physics\\
Lehigh University\\
Bethlehem, Pennsylvania 18015,\\
Jorge Vi\~nals\\
Supercomputer Computations Research
Institute, B-186\\
and\\
Department of Chemical Engineering, B-203\\
Florida State University\\
Tallahassee, Florida 32306-4052,\\
and\\
J.D. Gunton\\
Department of Physics\\
Lehigh University\\
Bethlehem, Pennsylvania 18015.}
\maketitle
\newpage

We present a numerical study of a model of pattern formation
following a convective instability in a non-Boussinesq
fluid. It is shown that many of the features observed
in convection experiments conducted on $CO_{2}$ gas
can be reproduced by using a
generalized two-dimensional Swift-Hohenberg
equation. The formation of hexagonal patterns, rolls and spirals
is studied, as well as the transitions and competition
among them. We also study nucleation and growth of hexagonal patterns
and find that the front velocity in this two dimensional model is consistent
with the prediction of marginal stability theory for one dimensional fronts.

\newpage

One of the most natural and intriguing behaviors of complex systems driven
far from thermal equilibrium is their ability to undergo symmetry breaking
instabilities that lead to the spontaneous formation of spatio-temporal
structures. An excellent example is the Rayleigh-B\'enard instability.
Much of the earlier experimental work has been restricted to
Oberbeck-Boussinesq type fluids, in which one observes various
configurations of
roll patterns. However, in a non-Oberbeck-Boussinesq system with,
for example,
temperature dependent transport coefficients, both roll and hexagonal
patterns can exist.
Very recently Bodenschatz et al. \cite{re:bo91} have performed
experiments on convection in $CO_{2}$ gas and studied
the existence of and transitions
between convective patterns exhibiting different symmetries.
They have observed the competition between a uniform conducting
state, a convective state with hexagonal symmetry, and
convecting patterns comprising parallel rolls.
In this letter we present the results of a numerical solution of
a two dimensional model equation for
the case of a large aspect ratio cylindrical cell near onset,
and qualitatively compare our results with the experiments of
Bodenschatz et al.

Our study is based on a numerical integration of a two dimensional
generalized Swift-Hohenberg equation \cite{re:sw77},
\begin{equation}
\label{eq:sh}
\frac{\partial \psi (\vec{r},t)}{\partial t} = \left[ \epsilon (\vec{r}) -
\left( \nabla^{2} + 1 \right)^{2}
- \psi^{2} \right] \psi + g_{2} \psi \nabla^{2} \psi + f(\vec{r}),
\end{equation}
with boundary conditions,
\begin{equation}
\psi|_{B} = \hat{n} \cdot \nabla \psi |_{B} = 0,
\end{equation}
where $\hat{n}$ is the unit normal to the boundary of the domain of
integration, $B$.
This equation with $g_{2}=0$ reduces to the Swift-Hohenberg equation and
has been extensively used to model convection in thin cells and near onset
\cite{re:co90,re:ma90,re:cr80,re:gr82}.
The scalar order parameter y
is related to the fluid temperature in the
mid-plane of the convective cell. The quantity $\epsilon$ is the reduced
Rayleigh number,
\begin{equation}
\epsilon = \frac{R}{R_{c}^{\infty}} -1,
\end{equation}
where
$R$ is the Rayleigh number and $R_{c}^{\infty}$ is the critical Rayleigh
number for an infinite system. A forcing field $f$ has been added
to mimic lateral sidewall forcing (for details, see
ref. \cite{re:xi91,re:vi91}).

We have derived a three mode amplitude equation from the generalized SH
equation in order to both estimate the threshold values of $\epsilon$ that
separate regions in which roll and hexagonal configurations are stable,
and the
values of the parameters that enter the generalized SH equation in
terms of experimentally measurable quantities. From the experiments
\cite{re:bo91},
we find that $g_{2} \approx 0.35$. This is the value that we have used
throughout our calculations. The value of $\epsilon$ used in the
numerical simulation is related to the real
experimental value $\epsilon^{exp}$ in ref. \cite{re:bo91}
by $\epsilon^{exp} = 0.3594 \epsilon$. The values of $\epsilon$ for
which hexagonal and roll patterns are stable can be obtained from the
three mode amplitude equation. For our value of $g_{2}$ we find that
hexagons are stable for $0 < \epsilon < \epsilon_{r}$, with
$\epsilon_{r} = 0.163$. Hexagonal patterns and rolls coexist for
$\epsilon_{r} < \epsilon < \epsilon_{b}$, with $\epsilon_{b} = 0.65$.
For $\epsilon > \epsilon_{b}$, only rolls are stable.
We have studied two different forms for the quandratic term in Eq.
(\ref{eq:sh}), $\psi \nabla^{2} \psi$ and  $\psi^{2}$.
We have found that both models give essentially the same results. In this
paper, we report on the results that correspond to $\psi \nabla^{2} \psi$
to study any non-variational effects on the hexagon to roll
transition. We have neglected the effect of mean flow
\cite{re:si81,re:ma83}
and have concentrated on the qualitative features of pattern formation
as described by Eq. (1). However, mean flow effects
are needed to describe, for example, a rotating spiral pattern, as shown
by Bestehorn et al. recently \cite{re:be92}.
In the following we report the results of our calculations.

i) Nucleation of a pattern with hexagonal symmetry. We consider as
initial
condition, $\psi( \vec{r},t=0)$, a Gaussian random
variable with zero mean and variance $10^{-6}$.
The forcing field $f(\vec{r})=0$
simply because there is no influence of lateral boundaries
before the nucleated pattern reaches the boundary.
We numerically solve Eq. (\ref{eq:sh}) in a square domain of side $L =
120 \pi $ (in our dimensionless variables, this corresponds to an aspect
ratio $\Gamma = L/ \pi = 120$).
The differential equation is discretized on a square grid with
$512 \times 512$ nodes.
We take $\epsilon = 0.01$
except in a small square region near the center of the cell (of size $16
\times 16$ nodes) where $\epsilon = 0.055$.
This space dependent $\epsilon$ models a small localized inhomogeneity
in one of the cell plates. The temporal evolution of the
pattern is shown in Fig. 1a.
It presents an early transient behavior during which a local convective
region with hexagonal symmetry has just nucleated.  Six fronts
of rolls are traveling away from the hexagonal
patch located at the center.
As they propagate further into the conduction region, they spread the
convective region with hexagonal symmetry.
This is remarkably similar to the experimental observations of
Bodenschatz et al. \cite{re:bo91}.
It is worth pointing out, however, that the shape of the
envelope of rolls is triangular in the experiment, whereas it
appears to be rectangular in our numerical results.
This difference may be attributable to a shortcoming of our model
equation. It is also possible, but probably not likely, that we have not
used a sufficiently small value of epsilon.
We have also estimated the speed of
propagation of the front that separates the hexagonal pattern and the
uniform state. This speed, at the center of the planar sides, and
along their normal direction, is constant in time and equals
$v_{\perp} = 0.37$.
The value given by marginal stability theory for the one dimensional
Swift-Hohenberg equation ($g_{2}=0$) is $v_{MS} = 0.397$
\cite{re:de83,re:sa88}. To our knowledge this front velocity
has not yet been experimentally measured.
The observation of nucleation and growth is especially interesting
since it provides an example of
competition among different symmetries, i.e. a uniform conduction state
as the background state, a region of hexagonal symmetry being nucleated and
rolls in the front region separating the two. This situation is also
interesting from the point of view of pattern selection
during front propagation in dimensions higher than one.

ii) Formation of patterns with hexagonal symmetry by sidewall forcing.
We have considered a circular cell of radius $R = 30 \pi$, which
corresponds to an aspect ratio of $\Gamma =
2R / \pi = 60$. A square grid with
$N^{2}$ nodes has been used with spacing $\Delta x = \Delta y = 60
\pi$/N, and $N=256$. We approximate the boundary conditions on $\psi$ by
taking $\psi (\vec{r},t) = 0$ for $\| \vec{r} \| \ge R$, where $\vec{r}$
is the location of a node with respect to the center of the domain of
integration. The initial condition $\psi (\vec{r},t=0)$ is a random
variable, gaussianly distributed with zero mean and variance $10^{-2}$.
In this case $\epsilon = 0.1$, and $f=0$ everywhere except on the nodes
adjacent to the boundary. Two values of $f$ have been studied: $f = 0.1$
and $f = 0.5$. They will be referred to as weak and strong forcing
respectively.
Figure 2a presents a typical configuration for
the case of weak sidewall forcing. Two ordered domains
appear in the cell with different orientation and
meet at an abrupt angle, creating the analog of a grain boundary
near a diameter of the cell.
On the contrary, we find that
a hexagonal pattern with concentric orientation
is induced spontaneously by the presence of strong
sidewall forcing, as shown in Fig. 2b.

iii) Transition between hexagons and rolls.
We use as initial condition the configuration shown in Fig. 2b,
with exactly the same
forcing field $f$ but take $\epsilon = 0.3$.
Figure 3 shows a sequence of configurations during the
hexagon to roll transition.
How defects mediate the transition can also be seen in Fig. 3.
They glide towards each
other, and invade nearby regions of hexagonal pattern to create a region
of rolls that spreads
through the cell as the transition proceeds. This is also very close to the
experimental observation, except in that
Bodenschatz et al. \cite{re:bo91} found that there is a tendency towards
the bending of rolls and the formation of spirals during the
hexagon to roll transition. This tendency does not appear in our numerical
results. This may be due to our neglect of mean flow effects in our
model.

iv) Stable spiral pattern. A one-armed spiral pattern is shown in
Fig. 4. We start with random initial
conditions in which $\psi(\vec{r},t=0)$ is a Gaussian random
variable with with zero mean and variance $10^{-8}$.
The forcing field at the boundary is $f = 0.1$ and $\epsilon = 0.2$.
The cylindrical cell has an aspect ratio of 60 and we have used
$256 \times 256$ grid nodes. We have observed that whenever a defect
appears, the neighboring rolls start to buckle and
a clear distortion occurs. This deformation develops, ending in
a stable spiral pattern. However, we are unable to see any rotation of
the spiral, presumably because of the lack of a mean flow term, as stated
earlier.

In summary, we have investigated the question of
pattern formation in a model of
convection in a non-Boussinesq fluid that allows stationary patterns of
various symmetries.
%
%
  Because the generalized Swift-Hohenberg equation that we have used
  is related to a non-Boussinesq fluid near the onset of convection,
%
%
%
we have been able to investigate
several interesting issues relevant to recent experimental observations
in $CO_{2}$ gas.
We have studied nucleation of hexagonal patterns, the
formation of patterns with hexagonal symmetry,
the transition from patterns with hexagonal symmetry to rolls and the formation
of a stable spiral pattern. Our numerical calculations have further
illustrated the strong influence of sidewall forcing in pattern formation.
Approximate boundaries in parameter space separating regions in which
patterns of the various symmetries occur, as well as a quantitative study of
front propagation will be given in future work.

We wish to thank E. Bodenschatz, G. Ahlers and D. Cannell for suggesting
the numerical investigation of the generalized Swift-Hohenberg equation,
and them and P. Hohenberg for many stimulating
conversations and comments. This work was supported in part
by the National Science Foundation under Grant No. DMR-9100245.
This work is also supported in part by the Supercomputer Computations
Research Institute, which is partially funded by the U.S. Department of Energy
contract No. DE-FC05-85ER25000. The calculations reported
here have been performed on the 64k-node Connection Machine at the
Supercomputer Computations Research Institute and on the Cray Y-MP at
the Pittsburgh Supercomputing Center.

\newpage

{\bf Figure captions}

Figure 1. Nucleation of a pattern with hexagonal symmetry
in a square cell with aspect ratio $\Gamma=120$. The values of the
parameters used are $g_{2} = 0.35, \epsilon = 0.01$ and $f=0$. In a
small square region at the center of the cell, $\epsilon = 0.055$. The
time shown is $t = 611$ and dark (white) areas represent regions in
which $\psi$ is positive (negative).

Figure 2. Hexagonal pattern obtained in a
cylindrical cell with aspect ratio $\Gamma=60$. The values of the
parameters used are $g_{2} = 0.35$ and $\epsilon = 0.3$. A non-zero
forcing localized at the boundary has been used. For weak forcing (a),
$f = 0.1$,
two domains of hexagonal symmetry with different orientation appear
that meet at a grain boundary.
For strong forcing (b), $f=0.5$, a uniform hexagonal pattern is
obtained at long times.

Figure 3. We observe a transition from hexagons to rolls by changing
$\epsilon$ from $\epsilon=0.1$
to $\epsilon=0.3$, in a cylindrical cell with an aspect
ratio $\Gamma=60$. The initial condition in a uniform hexagonal pattern
like the one shown in Fig. 2b. Two different times, $t=1712$ (a), and
$t=1819$ (b) are shown. The rolls appearing
near the defects and grain boundaries propagate along the axes which
adjoin the defects.

Figure 4. Stationary spiral obtained in a cylindrical cell with
aspect ratio $\Gamma=60$, with $g_{2} = 0.35, \epsilon = 0.2$ and
$f=0.1$. Concentric rolls are deformed
by the nucleation of defects which ultimately lead to a spiral pattern.

\end{document}